\documentstyle[12pt]{article}
\begin{document}
\newpage
\begin{center}
\Large {{\bf Magnetic Charge of the Stark States of Hydrogen Atoms}}
\end{center}
\medskip
\centerline{T. PRADHAN }
\centerline{\small \em
Institute of Physics, Bhubaneswar 751 005,Orissa India}
 \centerline{e-mail: pradhan@iopb.res.in }
\smallskip
 \centerline{PACS : 1480, Magnetic Monopoles}
\smallskip  
\centerline {{\bf\underline {Abstract}}}

It is conjectured that Stark states of excited hydrogen atom posses magnetic charge for which the quantum mechanical operator is 
$${\cal G}_{op} = {e\over \hbar} (\vec\sigma\cdot\vec A)$$
where $\vec A$ is the Runge-Lenz vector. The expectation value $g$ of this operator for Stark states is found to be 
$$ g = e(n_1-n_2)$$
which obeys a Dirac-Saha type quantization formula
$${eg\over c} = (n_1-n_2)\alpha$$
where $\alpha$ is the fine structure constant and $n_1$ and $n_2$ are parabolic quantum numbers. An experimental arrangement is outlined to test this conjecture.
\vfill
\eject
Magnetic charge density $\rho(\vec r)$ defined by 
$$\rho (\vec r) = \vec\nabla\cdot\vec B(\vec r)\eqno{(1)}$$
is a pseudoscalar. The quantum mechanical operator for it will therefore have vanishing expectation value for states with definite parity. However, for states with mixed parity, such as the degenerate states of hydrogen atom, this operator can have non-vanishing expectation value as is the case of electric dipole moment operator where the states are the Stark states. These are the particular combination of degenerate states of H-atom placed in an uniform external static electric field.

The three pseudosclar operators $(\vec L\cdot\vec A), (\vec L\cdot\vec p)$ and $(\vec L\cdot\vec r)$ of spinless hydrogen atom that can be costructed from $\vec r,\vec L,  \ \vec p$ and the Runge-Lenz vector $\vec A$ happen to be zero. However, if electron spin is taken into account, we can have three non-vanishing pseudoscalars which are $(\vec\sigma\cdot\vec A), \ (\vec\sigma\cdot\vec p)$ and $(\vec\sigma\cdot\vec r)$. The second of these is $\mid\vec p\mid$ times helicity. Out of the rest two, we prefer the first one since it is time-independent which is an essential requirement for the magnetic charge. Thus our choice, with correct dimensions, is
$${\cal G}_{op}= {e\over \hbar} \vec\sigma\cdot\vec A\eqno{(2)}$$

The Stark states are characterized by $\Psi_{n n_1n_2m}(\vec r)$ where parabolic quantum numbers $n, n_1,n_2,m$ are constrained by the relation
$$n = n_1+n_2+\mid m\mid+1\eqno{(3)}$$
One more state representing the spin orientation is required. We shall denote these as $\alpha$ and $\beta$ 
$$\alpha =\pmatrix{1\cr 0\cr }, \ \beta =  \pmatrix{0\cr 1\cr} \eqno{(4)}$$
which are spinors for spin up and down respectively. The total wavefunction is a product of $\Psi_{n,n_1n_2m}(\vec r)$and $\alpha$ or $\beta$ which has to be used to obtain the expectation value of $\vec \sigma \cdot\vec A$:
$$  <\vec\sigma\cdot\vec A >_{\alpha} = + < nn_1n_2\mid A_3\mid n n_1n_2>\eqno{(5)}$$
for spin up and 
$$  <\vec\sigma\cdot\vec A >_{\beta} = - < nn_1n_2\mid A_3\mid n n_1n_2>\eqno{(6)}$$
for spin down, since
$$\matrix{<\alpha\mid\sigma_1\mid\alpha > & = \alpha\mid\sigma_2\mid\alpha > =0\cr
<\beta\mid\sigma_1\mid\beta > & = \beta\mid\sigma_2\mid\beta > =0\cr
<\alpha\mid\sigma_3\mid\alpha > & =   +1\cr
 <\beta\mid\sigma_3\mid\beta > & = -1\cr}\eqno{(7)}$$
It may be noted that the Runge-Lenz vector $\vec A$ has to be properly symmetrized for hermiticity and normalized so as to satisfy the commutation relation
$$ [ A_1 A_2] = i L_3\eqno{(8)}$$
The expectation value for the third component such a Runge-Lenz vector is found to be
$$<n,n_1,n_2\mid A_3\mid n,n_1n_2 > = (n_1-n_2)\hbar  \eqno{(9)}$$
which on substitution in (2) and subsequent use of (5) and (6) gives for spin up and down.
$$g^{(\pm)} = \pm (n_1-n_2) e  \eqno{(10)}$$
The magnetic charge therefore  satisfies 
$${eg^{(\pm)}\over \hbar c} = \pm (n_1-n_2)\ \alpha \, \ \ \ \alpha ={e^2\over \hbar c}\eqno{(11)}$$
which is a Dirac-Saha type quantization formula.

That the H-atom Stark states have magnetic charge can be experimentally tested by exciting a beam of hydrogen atoms by appropriate laser beams to the state n=2 and passing the beam through a chamber in which they are subjected to an external uniform static electric field. The field automatically arranges the states into Stark states. If our conjecture of their possessing magnetic charge is correct then the two equal opposite magnetic charges will get deflected from their path by the action of the electric field just as equal and opposite electric charges do in an external static uniform magnetic charge. The deflected beams can be tested for magnetic charge  by passing them through SQUID detectors. The spins of the deflected beams can be manipulated appropriately by an external magnetic field.

Numerous experimental searches [2] have been undertaken to detect magnetic charge, and there seems to be some evidence for them in hadronic spectra [3] as well as in iron aerosols [4]. If experiment as suggested in this communication succeeds in finding magnetic charge in H-atom, it will be ample source of magnetic charge with which many experiments can be conducted which will reveal many interesting aspects of magnetism and  even have  technological applications.
\vfill
\eject
 \begin{enumerate} 
\centerline { References}
\vskip .5 cm
\item   P.A.M. Dirac, Proc. Roy. Soc. {\bf A133} 60 (1931), Phys. Rev. {\bf 74}
 817 (1948)\\
M. N. Saha, Ind. J. Phys. {\bf 10} 141 (1930) , Phys. Rev. {\bf 75} 1968 (1949)
\item G.Giacomelli et. al. Magnetic Monopole Bibliography DFUB 200-9, Bologna, May 2000
\item D. Akers, Int. J. of Theorem. Phys. {\bf 33} 1817 (1994).
\item V.F. Mikhalilov, Phys. Lett. {\bf 130B} 31    (1983)\\
J. Phys. {\bf A18} L903 (1985), {\bf A24} 53 (1991). 
\end {enumerate}
\end{document}